\documentclass[twocolumn,aps,prl] {revtex4} 
\usepackage{graphicx}
\begin{document}
\pagestyle{empty} 
\title{Rubber friction on (apparently) smooth lubricated surfaces}
\author{M. Mofidi, B. Prakash}
\affiliation{Division of Machine Elements, Lule\aa \ University of Technology, 
Lule\aa \ SE-97187, Sweden}
\author{B.N.J. Persson}
\affiliation{IFF, FZ-J\"{u}lich, 52425 J\"{u}lich, Germany}
\author{O. Albohr}
\affiliation{Pirelli Deutschland AG, 64733 H\"ochst/Odenwald, Postfach 1120, Germany}

\begin{abstract}
We study rubber sliding friction on hard lubricated surfaces. We show that even if
the hard surface appears smooth
to the naked eye, it may exhibit short wavelength roughness, which may give the dominant 
contribution to rubber friction. 
That is, the observed sliding friction 
is mainly due to the viscoelastic deformations of the
rubber by the substrate surface asperities. 
The presented results are of great importance for rubber sealing 
and other rubber applications involving (apparently) smooth surfaces.
\end{abstract}
\maketitle


{\bf 1. Introduction}

Rubber friction on smooth surfaces is a topic of great practical importance,
e.g., for rubber sealing, wiper blades or for the contact between a tire and the
metal rim\cite{Moore}. For perfectly smooth surfaces rubber friction is believed 
to be due to periodic cycles of
pinning, elastic deformation, and rapid slip of rubber molecules\cite{Schall1,Cher,Urbach} or,
more likely, small patches\cite{PerssonVolo} of the rubber at the sliding interface.
In a recent publication, Vorvolakos and Chaudhury\cite{Cha} 
(see also Ref. \cite{Caroli,Grosch}) have studied rubber 
friction for a silicone elastomer
sliding on extremely smooth Si wafer, with the root-mean-square
roughness $\approx 0.5 \ {\rm nm}$, covered by inert self-assembled monolayer films. 
The observed friction as a function of the sliding velocity exhibit a bell-like shape 
as expected from theory\cite{Schall1,PerssonVolo}.
However, a surface which appears smooth to the naked eye may exhibit
strong surface roughness at short length scales, e.g., at the 
micrometer and nanometer length scale. This is true even for highly
polished surfaces which may appear perfectly smooth to the naked eye. When
a rubber block slides on a hard surface with surface roughness, a large contribution to
the friction force may arise from the time-dependent, 
substrate asperity-induced deformations of the rubber surface. That is, during sliding 
the substrate asperities give rise to pulsating deformations of the rubber, which
will result in energy dissipation because of the internal friction of the rubber. 
This is believed to be the major contribution to the tire-road friction\cite{P8,Heinrich}. In this
paper we will show that the roughness of a highly polished steel surface may also give
the dominant contribution to the friction, even for lubricated surfaces.
This result is very important for rubber sealing applications\cite{Salant}, in particular at low sliding
velocities and low temperatures.

\vskip 0.3cm

{\bf 2. Rubber friction: experimental results}

\begin{figure}
\includegraphics[width=0.5\textwidth,angle=0]{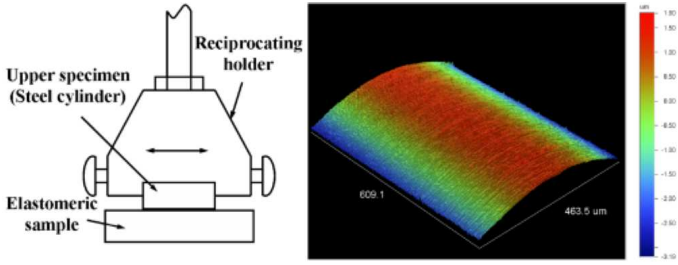}
\caption{\label{pic11}
Test configuration for friction studies under reciprocating sliding conditions.
}
\end{figure}

Friction tests have been carried out using a reciprocating tribometer where a steel cylinder  (diameter
$D=1.5 \ {\rm cm}$ and length $L=2.2 \ {\rm cm}$) is squeezed against
the substrate (rubber block, thickness $4 \ {\rm mm}$), see Fig. \ref{pic11}. 
The steel cylinder performs longitudinal oscillations against
the rubber block with a stroke $a=1 \ {\rm mm}$ and frequency $f=50 \ {\rm Hz}$.
This gives the average slip velocity $v \approx 0.1 \ {\rm m/s}$.
The rubber specimens [acronitrile butadiene rubber (NBR)]
have been washed in industrial petroleum for 3 minutes by using an ultrasonic 
cleaner and then dried for 10 minutes. The rubber surface has the root-mean-square roughness
$\approx 0.4 \ {\rm \mu m}$, and has parallel grooves caused during molding of elastomer
sheets in steel mold. The steel cylinder has a root-mean-square roughness of 
$\approx 0.1 \ {\rm \mu m}$.  

\begin{figure}
\includegraphics[width=0.45\textwidth,angle=0]{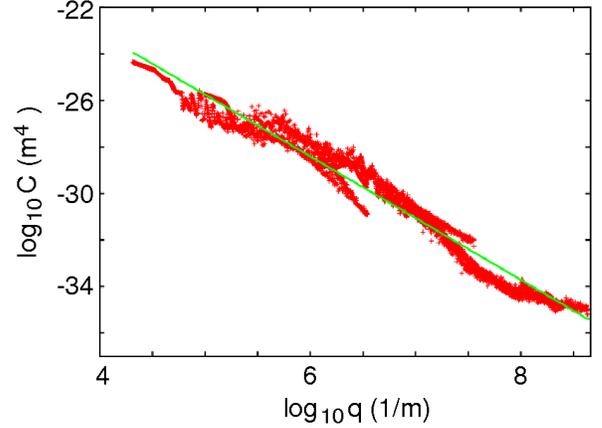}
\caption{\label{logq.logC}
The power spectrum of the surface roughness of the steel surface. 
The root-mean-square surface roughness is about $0.08 \ {\rm \mu m}$.
The straight line has a slope corresponding to the fractal dimension
$D_{\rm f} \approx 2.66$.
}
\end{figure}

Fig. \ref{logq.logC} shows 
the power spectrum of the surface roughness of the steel surface. 
The power spectrum is defined by\cite{PERS}
$$C(q)=\int d^2x \ \langle h({\bf x})h({\bf 0})\rangle e^{i{\bf q}\cdot {\bf x}}\eqno(1)$$
where $\langle .. \rangle$ stands for ensemble averaging. Here 
$h({\bf x})$ is the surface height at the point ${\bf x}$, where we have assumed 
$\langle h({\bf x})\rangle = 0$.
The surface height was measured over different surface areas using Atomic Force Microscopy
and an optical method, and Fig. \ref{logq.logC} was obtained from three different measurements
involving different resolution. 
The straight (green) line has a slope corresponding to the fractal dimension
$D_{\rm f} \approx 2.66$. In the calculations 
of the friction presented below we have 
used the this linear approximation and included the
surface roughness power spectra over the full wave vector range shown in the figure.
Thus the longest and the shortest wavelength roughness included in the analysis
is $\lambda_0=2 \pi /q_0 \approx 0.3 \ {\rm mm}$ and $\lambda_1 = 2 \pi /q_1 \approx 6 \ {\rm nm}$.

\begin{figure}
\includegraphics[width=0.48\textwidth,angle=0]{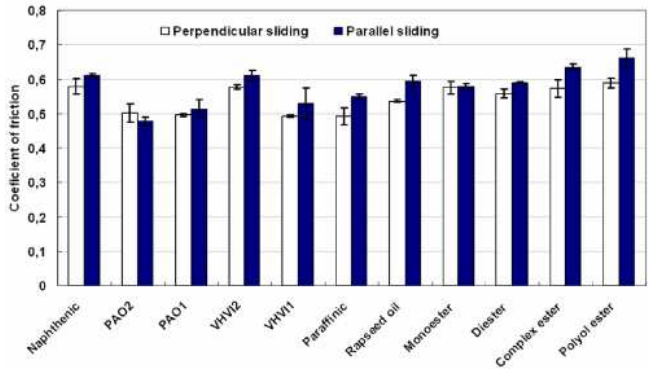}
\caption{\label{pic33}
Coefficient of friction of non-aged samples in different base oils.
For the load $F_{\rm N}=100 \ {\rm N}$.
}
\end{figure}

\begin{figure}
\includegraphics[width=0.48\textwidth,angle=0]{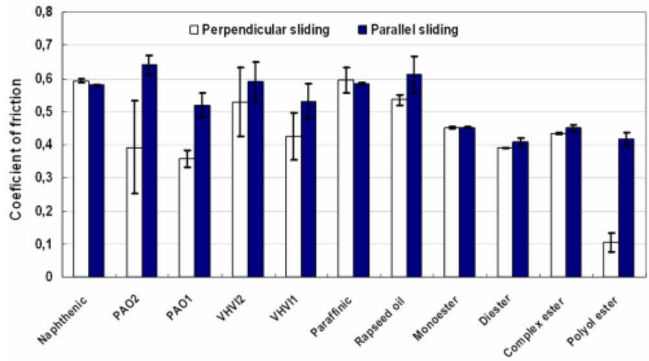}
\caption{\label{pic22}
Coefficient of friction of aged samples in different base oils.
For the load $F_{\rm N}=100 \ {\rm N}$.
}
\end{figure}

The experimental results presented in Fig.  \ref{pic33} and  \ref{pic22} 
were obtained for the load $F_{\rm N} = 100 \ {\rm N}$ and with a test duration
of 15 min. Since the oscillation stroke is very small (1 mm) one expects that most
of the oil is squeezed out from the steel cylinder -- rubber contact region. 

\begin{figure}
\includegraphics[width=0.45\textwidth,angle=0]{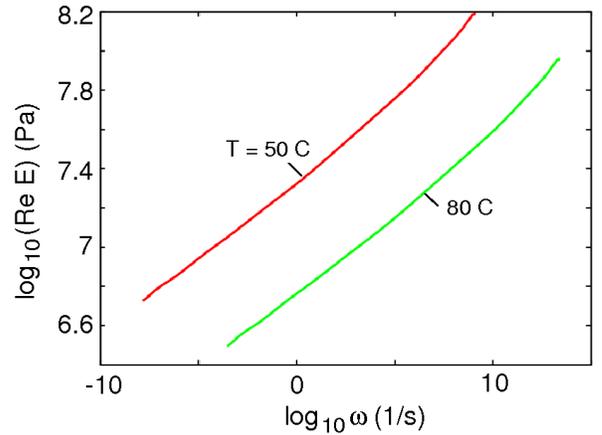}
\caption{\label{log.omeg.log.ReE.50.80}
The logarithm of the real part of the viscoelastic modulus as
a function of the logarithm of the frequency $\omega$ for the
temperatures $T= 50 \ ^\circ {\rm C}$ and $80 \ ^\circ {\rm C}$.
For acronitrile butadiene rubber.
}
\end{figure}

The viscoelastic modulus $E(\omega )$ has been measured
(using Eplexor 150) using a rectangular rubber block $5\times 2\times 30 \ {\rm mm}^3$.
The measurements were done in tension  
with $8\%$ of prestrain and $1.3\%$ of dynamic strain amplitude.
Fig. \ref{log.omeg.log.ReE.50.80} shows the 
logarithm of the real part of the viscoelastic modulus of the acronitrile butadiene rubber 
used in the present study, as
a function of the logarithm of the frequency $\omega$, for the
temperatures $T= 50 \ ^\circ {\rm C}$ and $80 \ ^\circ {\rm C}$.

\begin{figure}
\includegraphics[width=0.3\textwidth,angle=0]{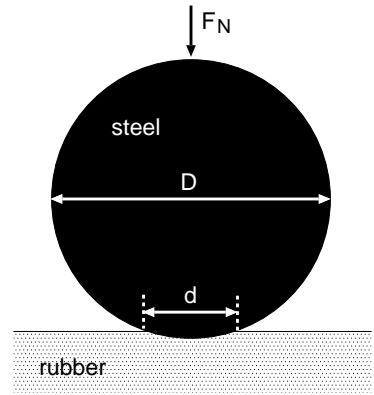}
\caption{\label{steelball}
Steel cylinder squeezed in contact to a rubber substrate.
}
\end{figure}

The diameter $d$ of the contact region between the steel cylinder and the rubber substrate
can be estimated using the Hertz contact theory for bodies with cylinder geometry,
see Fig. \ref{steelball}. For elastic solids, the
diameter $d$ of the contact area is given by\cite{Johnson}
$$d=2 \left ({2F_{\rm N} D \over \pi L E^*}\right )^{1/2},\eqno(2)$$
where $E^*=E/(1-\nu^2)$ (where $E$ is the Young modulus and $\nu$ the Poisson
ratio).
The average pressure in the contact region is
$$\bar p = {1\over 2} \left ({\pi F_{\rm N} E^* \over 2 L D}\right )^{1/2}.\eqno(3)$$
For $F_{\rm N}=100 \ {\rm N}$ and for $T\approx 50 \ ^\circ {\rm C}$ we have
(see Fig. \ref{log.omeg.log.ReE.50.80}) $E^* \approx 10 \ {\rm MPa}$ 
(where we have assumed the frequency $\omega \approx 10^{-3} \ {\rm s}^{-1}$, corresponding to the
contact time $\sim 1000 \ {\rm s}$) giving
$d \approx 0.4 \ {\rm cm}$ and $\bar p \approx 1 \ {\rm MPa}$.

\begin{figure}
\includegraphics[width=0.48\textwidth,angle=0]{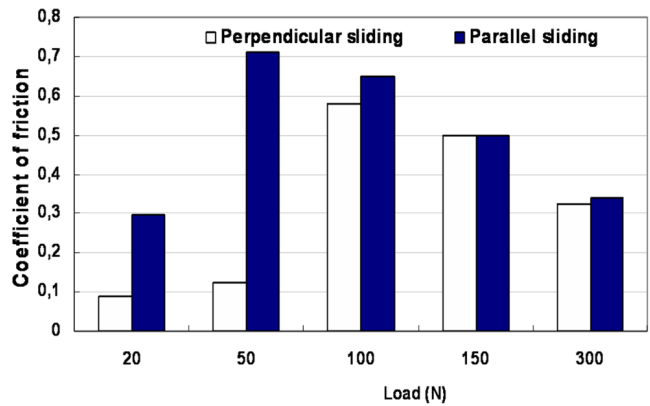}
\caption{\label{bild1}
Friction coefficient as a function of load at the background temperature $T= 25 \ ^\circ {\rm C}$.
}
\end{figure}

\begin{figure}
\includegraphics[width=0.48\textwidth,angle=0]{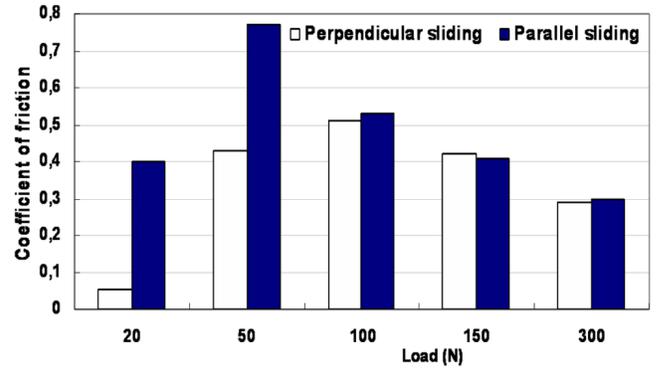}
\caption{\label{bild2}
Friction coefficient as a function of load at the background temperature $T= 40 \ ^\circ {\rm C}$.
}
\end{figure}

\begin{figure}
\includegraphics[width=0.48\textwidth,angle=0]{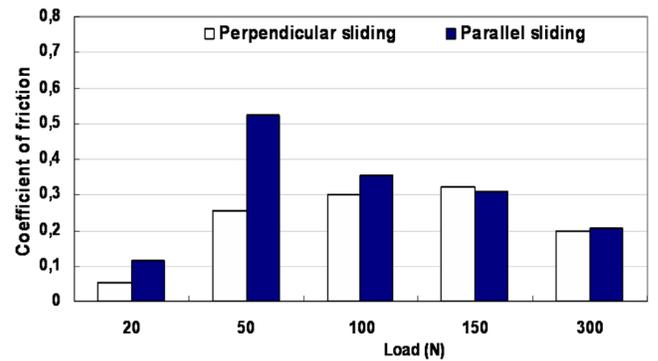}
\caption{\label{bild3}
Friction coefficient as a function of load at the background temperature $T= 80 \ ^\circ {\rm C}$.
}
\end{figure}

\begin{figure}
\includegraphics[width=0.48\textwidth,angle=0]{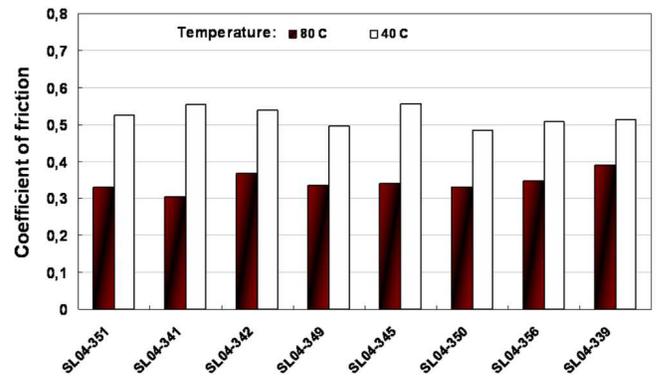}
\caption{\label{bild4}
Friction coefficient for one base oil with several different additives and for $T= 40 \ ^\circ {\rm C}$ 
and $80 \ ^\circ {\rm C}$.
For the load $F_{\rm N}=100 \ {\rm N}$.
}
\end{figure}

Fig. \ref{pic33} shows the measured friction coefficients for
the steel cylinder sliding against non-aged rubber in 11 different lubrication
oils with very different viscosities. Thus, for example, the PA01 and PA02 oils
have the viscosities (at $T=40 \ ^\circ {\rm C}$) $4.4\times 10^{-3}$ and 
$22.8\times 10^{-3} \ {\rm Pa \ s}$,
respectively. In spite of the large difference in viscosities, the rubber friction coefficients
are nearly equal. This indicates that the rubber friction is not (mainly) due to shearing a thin
viscous layer, but due to the internal friction of the rubber (see below).

Fig. \ref{pic22} shows the  measured friction coefficients for aged rubber. The aged rubber
samples were prepared by immersing them in different base fluids at $T = 125 \ ^\circ {\rm C}$
for one week. 
NBR rubber has polar nitrile groups and non-polar oils such as naphthenic have nearly no effect on
the properties of NBR rubber, and this explained why rubber aged in naphthenic exhibits nearly
the same friction as for non-aged NBR rubber (compare Fig. \ref{pic33} with \ref{pic22}).
However, oils with polar groups, e.g. polyol ester, will
diffuse into the rubber which may reduce the internal
friction of the rubber. In addition, when the rubber block is squeezed against the substrate, 
oil may be squeezed
out from the rubber matrix, giving a thicker oil film at the 
interface and thus lower the friction 
(a similar effect is believed to contribute to the
extremely low friction exhibited by human joints\cite{PerssonBook}).
We believe that both effects may contribute to why NBR rubber
aged in polyol ester exhibits much
smaller friction than the non-aged rubber. 

Figures \ref{bild1}, \ref{bild2} and \ref{bild3} show the friction coefficient for different loads 
and temperatures. Here the temperature refers to the background temperature, which was varied by contacting
the back-side of the rubber block to a metal block with the given temperature. (The temperature
in the sliding contact is not known, but will be higher due to the frictional heating.)
Note that as the temperature increases the friction decreases. This cannot result from the
change in viscosity of the lubricant oil since we already know from above that the lubricant viscosity
has a negligible influence on the friction, at least for the squeezing force 
$F_{\rm N} = 100 \ {\rm N}$, see Fig. \ref{pic33}.
However, we will show in Sec. 3 that the temperature dependence of the sliding friction 
can be understood from the
temperature dependence of the internal friction of the rubber. Thus, when the temperature increases
the rubber becomes more elastic (less viscous) and the internal friction decreases.

The dependence of the rubber friction on the load can be understood as follows. For very small load
($F_{\rm N} = 20 \ {\rm N}$) the average pressure in the contact area [see Eq. (3)] 
is relative low and the grooves
on the rubber surface will not be (fully) elastically flattened, and will trap lubricant oil, which may 
be pulled into the contact area during each oscillation.
This will result in an oil film which is thick enough to reduce the rubber--steel asperity
contact and hence lower the viscoelastic contribution to the friction. This drag of lubricant fluid
into the contact area is particularly large when the oscillation direction is perpendicular to the 
grooves on the rubber surface\cite{Drag}, and this explains why the friction for small load is much lower for 
perpendicular sliding than parallel sliding. However, for high load ($F_{\rm N} \ge 100 \ {\rm N}$)
there is negligible difference between parallel and perpendicular sliding, indicating that
the lubricant has a negligible direct influence on the friction.

The drop in the friction for large load is most likely due to the increase in the temperature
caused by the frictional heating. This effect becomes more important as the load increases, and 
explains why the friction decreases for high load. At lower sliding velocity (or oscillation frequency)
the heating effects become less important (because of heat diffusion) and in this case one expects a
smaller drop in the friction coefficient with increasing load. We plan to test this prediction
experimentally.

Fig. \ref{bild4} shows the friction coefficients (for the load $F_{\rm N}=100 \ {\rm N}$) at $T=40 \
^\circ {\rm C}$ and $80 \ ^\circ {\rm C}$ for the same base oil but with different additives. As expected,
there is negligible dependence of the friction on the additives. The reason for this is the same as before:
the observed friction is mainly due to the internal friction of the rubber which does not 
change between the different experiments. That is, although 
the additives in the base oil may adsorb on the solid
surfaces and act as boundary lubricants, the result of the study above indicates
that such (mono) layers have negligible influence on the friction.

\vskip 0.3cm

{\bf 3. Rubber friction: theory}

We have calculated the dependence of the rubber friction on the sliding velocity and the temperature
using the theory presented in Ref. \cite{P8}. The theory assumes that the friction is 
entirely due to the viscoelastic deformation of the rubber, which
results from the pulsating deformations from the
substrate asperities. The only inputs in the calculations are the surface roughness power spectrum 
(see Fig. \ref{logq.logC}) and the rubber viscoelastic modulus. We have measured the viscoelastic modulus 
$E(\omega )$ of the rubber as a function of frequency (and temperature). In the calculations we do not
take into account the lubrication oil directly (but it influences the friction
indirectly by reducing (or removing) the
adhesion between the solid walls\cite{Bruno}).
We have assumed the nominal contact pressure of $1 \ {\rm MPa}$.

\begin{figure}
\includegraphics[width=0.45\textwidth,angle=0]{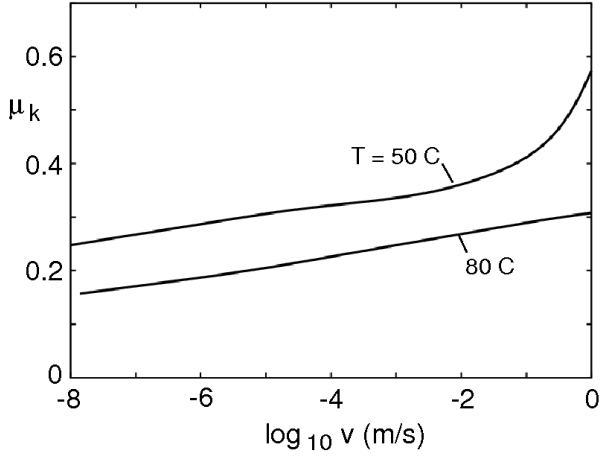}
\caption{\label{logv.mu}
The steady state kinetic friction coefficient calculated using the measured
surface roughness power spectrum (from Fig. \ref{logq.logC}) and the measured viscoelastic modulus
of the rubber. For the background temperatures $50^\circ {\rm C}$ and $80^\circ {\rm C}$, 
and the nominal squeezing pressure $p=1 \ {\rm
MPa}$.
}
\end{figure}

Fig. \ref{logv.mu} shows
the steady state kinetic friction coefficient calculated using the measured
surface roughness power spectrum (from Fig. \ref{logq.logC}) and the measured viscoelastic modulus
of the rubber. Results are presented for the background 
temperatures $50^\circ {\rm C}$ and $80^\circ {\rm C}$.
Note that the magnitude of the calculated friction coefficient at the sliding velocity
$\sim 0.1-1 \ {\rm m/s}$ is similar to what is observed experimentally, and also the temperature
dependence is in good agreement with the measurements (see Sec. 2). 

\begin{figure}
\includegraphics[width=0.45\textwidth,angle=0]{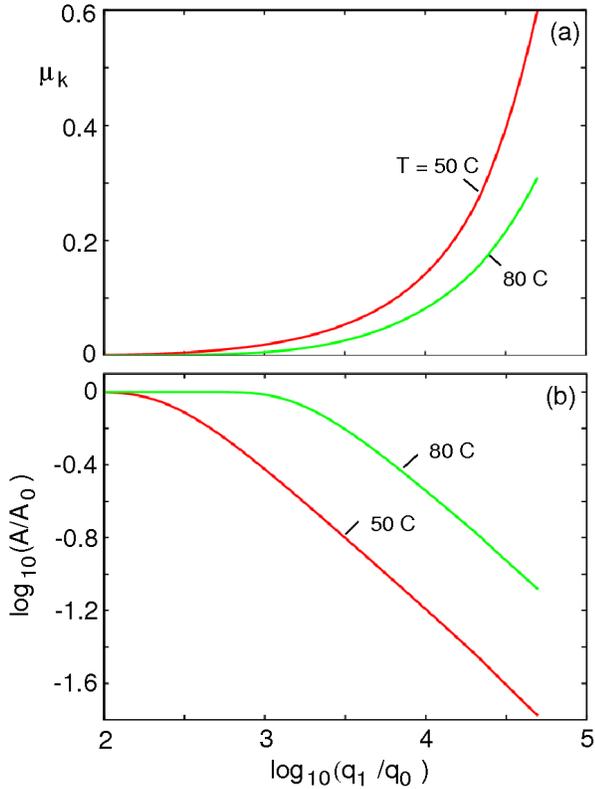}
\caption{\label{log.q.mu.area.40.60}
The friction coefficient $\mu_{\rm k}$ (a) and the logarithm of the (normalized) 
contact area $A/A_0$ (b),
as a function of the logarithm of the 
large-wavevector cut-off $q_1$ [in units of the low-wavevector cut-off $q_0$]. 
In the calculations we  
only include surface roughness with wavevectors $q_0 < q < q_1$.
Results are presented for two different temperatures 
$T= 50 \ ^\circ {\rm C}$ and $80 \ ^\circ {\rm C}$ and for the sliding velocity $v=1 \ {\rm m/s}$.
}
\end{figure}

In Fig. \ref{log.q.mu.area.40.60} we show (a)
the friction coefficient $\mu_{\rm k}$, and (b) the logarithm of the (normalized) 
contact area $A/A_0$,
as a function of the logarithm of the 
large-wavevector cut-off $q_1$ (in the calculations we  
only include surface roughness with wavevectors $q_0 < q < q_1$).
Results are presented for two different temperatures 
$T= 50 \ ^\circ {\rm C}$ and $80 \ ^\circ {\rm C}$ and for the sliding velocity $v=1 \ {\rm m/s}$.
The figure shows that the long-wavelength roughness gives a negligible contribution
to the friction. The reason for why only the short-wavelength roughness is
important in the present case is the large fractal dimension ($D_{\rm f} \approx 2.7$) of the steel surface,
which implies that the ratio between the amplitude and the wavelength of the surface roughness strongly
increases as the wavelength decreases\cite{amplratio}, 
and this makes the short-wavelength roughness much more important
than the long-wavelength roughness.

\vskip 0.3cm

{\bf 4. Squeeze-out}

We have argued above that the observed rubber friction can be explained as resulting from the viscoelastic
deformations of the rubber by the countersurface asperities. In this section we briefly address the
role of the lubrication oil. We first note that the oil will effectively eliminate the
adhesive interaction between the rubber and the countersurface\cite{Bruno}. Most of the oil
will be squeezed out from the steel-rubber contact area, but a molecular thin layer may 
remain even after long squeezing time.

\begin{figure}
\includegraphics[width=0.45\textwidth,angle=0]{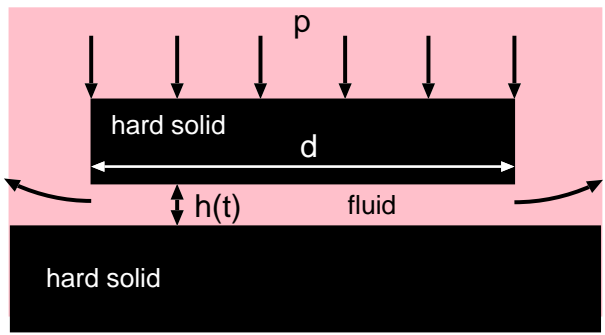}
\caption{\label{squeezeout}
A block squeezed against a substrate in a fluid. The separation
between the bottom surface  of the block and the top surface of the substrate
is denoted by $h(t)$. 
}
\end{figure}

\begin{figure}
\includegraphics[width=0.45\textwidth,angle=0]{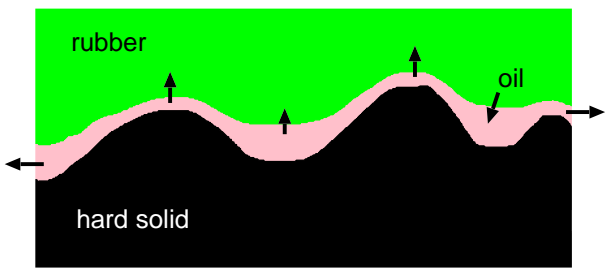}
\caption{\label{oilinter}
A rubber block squeezed against a substrate in an oil. The oil is
partly squeezed out at the external boundaries of the nominal contact area
and partly transfered to (or from) the rubber matrix. 
}
\end{figure}

Consider first a flat rigid rectangular block squeezed against a flat hard countersurface 
with the nominal (or average) pressure $p$ in a lubrication fluid with the viscosity $\eta$.
The separation between the surfaces after the time $t$ is 
(see Fig. \ref{squeezeout})\cite{PerssonBook}
$$h(t) \approx ( \eta /2 p t)^{1/2} d.\eqno(4)$$
Here $d$ is the width of the bottom surface of the block and we assume that $d<<L$, where $L$ is the length
of the bottom surface of the block. 
With $d\approx 0.4 \ {\rm cm}$, $p\approx 1 \ {\rm MPa}$ and with $t= 1000 \ {\rm s}$
we get with the typical viscosity 
$\eta \approx 0.01 \ {\rm Pa \ s}$, $h(t) \approx 4 \ {\rm nm}$.
For surfaces with roughness the squeeze-out from asperity contact regions is even faster, but in this
case some liquid may get ``trapped'' in sealed off regions\cite{Mugele}. For non-aged rubber the trapped
islands may disappear because of diffusion of lubricant
oil into the rubber matrix, see Fig. \ref{oilinter}. The shear stress developed in a fluid film
with thickness $h$ is $\sigma= \eta v /h$. In the present case, if $v=0.1 \ {\rm m/s}$ and
$h=10 \ {\rm nm}$ we get $\sigma= 0.1 \ {\rm MPa}$ which would 
give a contribution to the friction coefficient of 
order $\sigma /p \approx 0.1$. However, the thickness of the oil film will be very non-uniform,
and in many regions (cavity regions) at the interface the 
film may be much thicker than $10 \ {\rm nm}$ (see below),
and shearing the lubricant film in these regions will give a negligible contribution
to the friction. In other regions, where the steel asperities make direct contact with the rubber, the local
squeezing pressure is much higher than the average pressure, and in these regions at most a few
monolayers of oil film will remain trapped. Nevertheless, since the region of direct wall-wall contact is only a
small fraction of the nominal contact area, the contribution to the friction from shearing the
confined thin layers appears to be negligible (see Sec. 2).

\begin{figure}
\includegraphics[width=0.45\textwidth,angle=0]{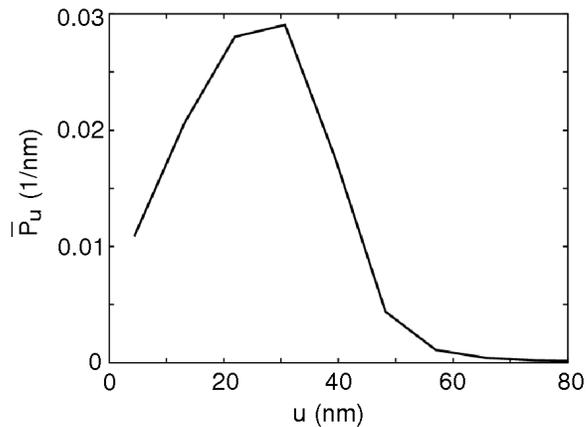}
\caption{\label{probability.distribution.for.u}
The calculated probability distribution $\bar P_u$ of surface separations $u$. 
}
\end{figure}

Fig. \ref{probability.distribution.for.u} shows the probability distribution $\bar P_u$ of surface
separations $u$. This function has been calculated as outlined in Ref. \cite{PerssonYang}. 
In the calculation we have assumed a rubber elastic modulus $E=100 \ {\rm MPa}$ which 
correspond to the temperature $T=40 \ ^\circ {\rm C}$ and the perturbing frequencies $\omega \approx 10^6
\ {\rm s}^{-1}$ (see Fig. \ref{logq.logC}), which is a typical perturbing frequency ($\omega = q v$) 
from surface roughness with wavevector 
$q = 10^7 \ {\rm m}^{-1}$ and sliding velocity $v=0.1 \ {\rm m/s}$.
In the calculation
we have neglected the direct influence of the lubrication oil, but it is accounted for indirectly by
neglecting the adhesive interaction between the rubber and the steel surface. Using $\bar P_u$
we can give a more accurate estimate of the contribution from the oil film to the shear stress. We get
the viscous shear stress
$$\sigma \approx \eta v \int_{u_{\rm c}}^\infty du \ {\bar P_u \over u}\eqno(5)$$
where $u_{\rm c}$ is a cut-off length of order $\sim 1 \ {\rm nm}$ since molecular thin lubrication
films cannot be described by the continuum theory of fluid mechanics\cite{Mugele}. 
We note that $\bar P_u$ has a 
delta function at the origin $u=0$, but in the present 
case this carry the weight $A(\zeta_1)/A_0 \approx 0.01$
and the contribution from the area of real contact to the friction force can be neglected.
Using the calculated $\bar P_u$ (see Fig. \ref{probability.distribution.for.u}),
and assuming $\eta=0.01 \ {\rm Pa \ s}$ and $v=0.1 \ {\rm m/s}$, Eq. (5) gives
$\sigma \approx 0.06 \ {\rm MPa}$ so the contribution to the friction from the lubricant film
is very small, of order $0.06$ (where we have assumed the normal stress $p= 1 \ {\rm MPa}$).
We note that this is likely to be an overestimation of the contribution of the
oil film to the friction coefficient, as the oil film may tend to slightly increase the 
separation between the walls, and also because we have not accounted for the roughness on
the rubber surface in the analysis.  

\vskip 0.3cm

{\bf 5. Discussion}

When a block of a viscoelastic solid, such as 
rubber, is sliding on a hard rough countersurface, the largest contribution
to the sliding friction is usually derived from the time-dependent deformations of the
rubber by the countersurface asperities. This is the case, for example, for the tire-road contact.
Here we have shown that even a highly polished countersurface, which may appear
mirror-smooth to the naked eye, may have enough surface roughness
at short length scale to give a large contribution to rubber friction. This result
has many important applications, e.g., in the context of rubber sealing. 

In many rubber sealing applications the rubber surface and the (lubricated) countersurface are squeezed together
for a long time between the slip events. Furthermore, during the onset (and stop) of sliding the slip
velocities will be very small. This may result in nearly complete squeeze-out of the lubricant film. 
Thus, at some point in time slip will occur between what is effectively 
unlubricated surfaces. This may result in high friction and large wear, and perhaps
failure of the seal with potentially serious consequences. 

Note that with respect to sliding friction there is an asymmetry between roughness on the 
countersurface and on the rubber block. Thus, only roughness on the hard countersurface will contribute to
the friction force. Roughness on the rubber surface may in fact {\it lower} the sliding friction by trapping
lubrication fluid. On the other hand, with respect to stationary contact mechanics, 
roughness on the two surfaces plays a similar role\cite{Johnson,SSR}.

There is an important difference between rubber friction on very rough surfaces, such as a road
surface, and rubber friction on smoother surfaces with only short-wavelength roughness. 
On very rough surfaces, as the magnification increases we observe smaller and smaller rubber-countersurface
asperity contact regions, and the local stress and temperature will rapidly increase until the limit
of strength of the rubber has been reached. For tread rubber in contact with road surfaces this 
limit is reached at the length scale (or resolution) 
$\lambda_{\rm c} \approx 1-10 \ {\rm \mu m}$, 
and at this length scale during slip strong wear processes occur.
The rubber friction on road surfaces can be explained by including the the viscoelastic
deformations of the rubber from road surface roughness
with wavelength down to $\lambda_{\rm c}$. 
On the other hand, for surfaces with mainly short-wavelength roughness, such as
the steel surface used in the present study, it may be necessary to include roughness with wavelength
down to the molecular length scale, e.g., the distance between cross links in the rubber 
which typically is of order a few nanometers. 
This may result in different wear mechanisms and wear rates than on
surfaces with large long wavelength roughness.

The results presented in this paper may also be relevant for the adhesion
and locomotion of some animals on rough substrates. Thus, some animals, such as grasshoppers and
tree frogs, have smooth attachment pads built from a (non-compact) material which is highly 
viscoelastic (like rubber)\cite{Good}. Furthermore, the toe pad-substrate contact region is wet (lubricated)
with a liquid injected into the contact area by the animal. The liquid viscosity, the nominal
squeezing pressure, and the size and shape of the contact area
differ from the lubricated rubber-substrate contact problem studied above, 
but some of the results presented above may nevertheless be relevant for the animal toe pad-substrate
interaction problem\cite{ToePad,Federle}.

\vskip 0.3cm

{\bf 6. Summary and conclusion}

We have presented a combined experimental-theoretical study of 
rubber sliding friction against hard lubricated 
surfaces. We have shown that even if
the hard surface appears smooth
to the naked eye, it may exhibit short wavelength roughness, which may give the dominant 
contribution to rubber friction. 
The presented results may be of great importance for rubber sealing 
and other rubber applications involving (apparently) smooth surfaces.


\begin{thebibliography}{999}
\bibitem{amplratio}
For a self affine fractal surface the ratio between the height $h(\lambda)$ and wavelength $\lambda$
of the surface roughness component with wave vector 
$q=2\pi /\lambda$ is $h/\lambda \sim \lambda^{2-D_{\rm f}}$.
Thus the larger the fractal dimension $D_{\rm f} > 2$, the faster the ratio 
$h/\lambda$ and will increase as the wavelength decreases, and this will tend to increase the
importance of the short wavelength roughness.
\end{thebibliography}
\end{document}